\documentclass[letter,
               keeplastbox,   % flushend option: not to un-indent last line in References
               ]{jacow}
\usepackage{pdfpages,multirow,ragged2e,mathtools} %

\ifboolexpr{bool{xetex} or bool{luatex}} % test for XeTeX/LuaTeX
 {}                                      % input encoding is utf8 by default
 {\usepackage[utf8]{inputenc}}           % switch to utf8

\usepackage[USenglish]{babel}

\usepackage{amsmath,amsfonts}
\usepackage{mathtools}
\usepackage{algorithm}
\usepackage{algpseudocode}
\usepackage{booktabs}
\usepackage[flushleft]{threeparttable}

\title{Adaptive Model-Based Reinforcement Learning for Orbit Feedback Control in NSLS-II Storage Ring}

\author{Z. Dong\thanks{Zeyu.Dong@stonybrook.edu}, Stony Brook University, Stony Brook, NY, USA\\
Y. Tian\thanks{ytian@bnl.gov}, Brookhaven National Laboratory, Upton, NY, USA\\
Y. Sun\thanks{yu.sun@sunriseaitech.com}, Sunrise Technology Inc., Stony Brook, NY, USA%
}

\DeclarePairedDelimiter{\norm}{\lVert}{\rVert}

\DeclareMathOperator*{\argmin}{arg\,min}
\def\sym#1{\ifmmode^{#1}\else\(^{#1}\)\fi}

\begin{document}
\maketitle
\begin{abstract}
% Abstract: The National Synchrotron Light Source II (NSLS-II) uses a highly stable electron beam to produce high-quality X-ray beams with high brightness and low-emittance synchrotron radiation. The traditional algorithm to stabilize the beam applies singular value decomposition (SVD) on the orbit response matrix to remove noise and extract actions. Supervised learning has been studied on NSLS-II storage ring stabilization and other accelerator facilities recently. Several problems, for example, machine status drifting, environment noise, and non-linear accelerator dynamics, remain unresolved in the SVD-based and supervised learning algorithms. To address these problems, we propose an adaptive training framework based on model-based reinforcement learning. This framework consists of two types of optimizations: trajectory optimization attempts to minimize the expected total reward in a differentiable environment, and online model optimization learns non-linear machine dynamics through the agent-environment interaction. Through online training, this framework tracks the internal status drifting in the electron beam ring. Simulation and real in-facility experiments on NSLS-II reveal that our method stabilizes the beam position and minimizes the alignment error, defined as the root mean square (RMS) error between adjusted beam positions and the reference position, down to ~1µm.
The National Synchrotron Light Source II (NSLS-II) uses highly stable electron beam to produce high-quality X-ray beams with high brightness and low-emittance synchrotron radiation.
The traditional algorithm to stabilize the beam applies singular value decomposition (SVD) on the orbit response matrix to remove noise and extract actions. 
Supervised learning has been studied on NSLS-II storage ring stabilization and other accelerator facilities recently. Several problems, for example, machine status drifting, environment noise, and non-linear accelerator dynamics, remain unresolved in the SVD-based and supervised learning algorithms.
To address these problems, we propose an adaptive training framework based on  model-based reinforcement learning.
This framework consists of two types of optimizations:
trajectory optimization attempts to minimize the expected total reward in a differentiable environment, and online model optimization learns non-linear machine dynamics through the agent-environment interaction.
Through online training, this framework tracks the internal status drifting in the electron beam ring.
Simulation and real in-facility experiments on NSLS-II reveal that our method stabilizes the beam position and minimizes the alignment error, defined as the root mean square (RMS) error between adjusted beam positions and the reference position, down to \textasciitilde\SI{1}{\micro\metre}.
\end{abstract}
%%%%%%%%%%%%%%%%%%%%%%%%%%%%%%%%%%%%%%%%%%%%%%%%%%%%%%%%%%%%%
%%%%%%%%%%%%%%%%%%%%%%%%%%%%%%%%%%%%%%%%%%%%%%%%%%%%%%%%%%%%%
\section{Introduction}
\begin{figure}[htb]
\centering
\includegraphics[width=.6\linewidth]{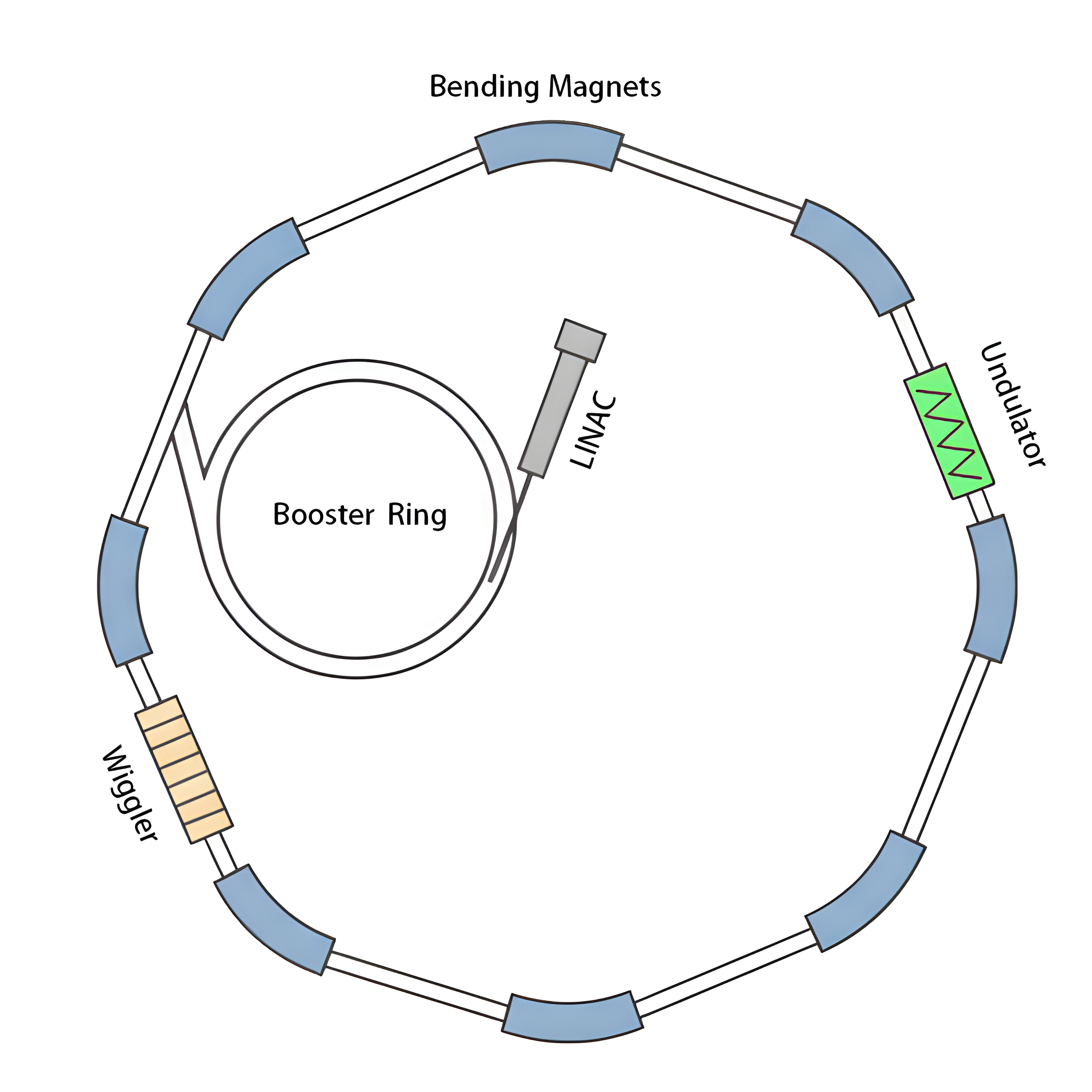}
\caption{An illustration of a storage ring.}\label{fig:beamline}
\end{figure}

NSLS-II is a third-generation storage ring producing synchrotron radiation through laser-electron interactions. 
Electrons are accelerated through a synchrotron and injected into the storage ring.
Low emittance in a light source facility requires stable electron beam orbit \cite{singh_nsls-ii_2013}. 
Figure~\ref{fig:beamline} shows a simplified electron orbit that remains in the beam position and emits X-ray radiation at multiple X-ray experiment locations. 

The beam operators rely on beam monitoring and control systems to interact with orbits. 
In each unit of the storage ring, beam position monitors  (BPMs) measure the relative position of the beam.
Each storage ring also includes corrector control to adjust beam dynamically. 
Ideally, the corrector currents in orbit controls are initialized at the beginning of experiments, and their initialization depends only on the design of the light source facility. 
In reality, noise and environmental change cause the beam to gradually drift away from the reference position. Orbit feedback control systems \cite{haga_global_1990} apply corrections to corrector controls and regains the beam reference position.

The orbit response matrix (ORM) reflects how BPMs respond to the correction. The matrix is static and belongs to the original machine design.
In practice, beam operators periodically measure the ORM by tuning  the beam close to the golden beam position and changing the current setting on one corrector one by one.
However, we cannot obtain precise measurement on the ORM because of the following reasons:
\begin{enumerate}
\item the machine dynamics drifts slightly over time, due to external environment influence, for example, hysteresis of the correctors \cite{choi_reproducibility_2016}, room temperature;
\item Nonlinear beam; the linear approximation of the ORM measured at the previous time incurs a large bias in modeling the beam system that is non-linear and evolves continuously during facility operation;   
\item Environment factors, including vibration and electronic noise, introduce measurement error (noises) in ORM.
\end{enumerate}

\begin{figure*}[htb]
\centerline{\includegraphics[width=\textwidth]{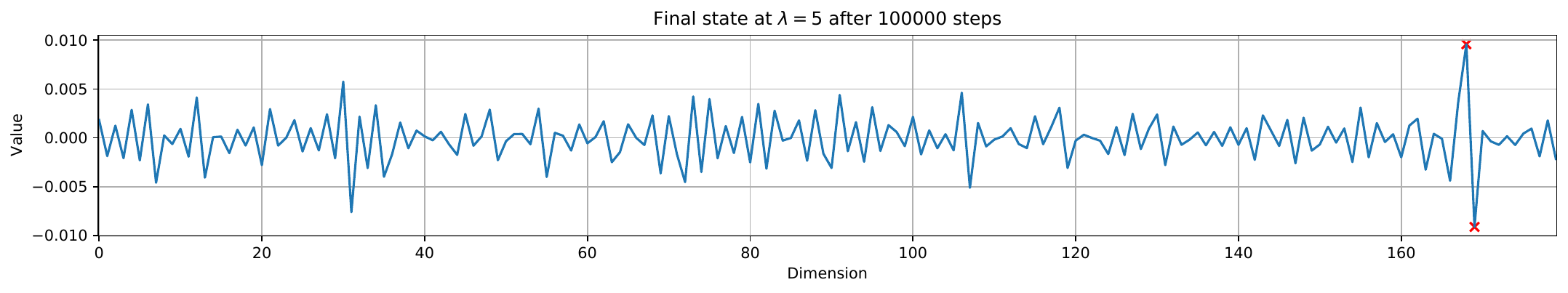}}
\caption{A degenerated final state after a long run. The dimensions that lose control are marked in red.}\label{fig:lambda-final}
\end{figure*}

Traditionally, SVD-based algorithms produce feedback signals by filtering out high-frequency components in ORMs \cite{corbett_algorithms_1994}.
Modern NSLS-II design for fast orbit feedback control combines SVD-based feedback control with other controllers, such as PID (proportional–integral–derivative), to ensure robust and stable beam orbits \cite{singh_nsls-ii_2015}.
However, this method has several limitations in practice.
Empirically, a large \(\lambda\) value leads to robust control to the ORM errors while generating biased RMS values.
On the other hand, because of the machine's internal drifting, the controller still has a degenerated performance, even with sufficiently large \(\lambda\).
Figure~\ref{fig:lambda-final} shows that after a long time, some dimensions will lose control even with very large \(\lambda\).

We seek an adaptive feedback control system to address this problem.
Reinforcement learning (RL) is a good solution for robots, self-driving, optimization and scheduling, and control systems.
With agent-environment interaction, the RL agent learns the behaviors of the orbit system and captures any machine drifting.
Multi-layer Neural networks (NN) in RL can be trained to model the non-linear dynamics of the orbit system.
However, high dimensional control with reinforcement learning is a challenging task.
Our feedback system in NSLS-II is high-dimensional and consists of 180 inputs for BPM measurements and 180 outputs of control systems.
Therefore,   we must use prior knowledge to model the target machine  and use the model to regularize RL training and overcome the curse of dimensionality.
%%%%%%%%%%%%%%%%%%%%%%%%%%%%%%%%%%%%%%%%%%%%%%%%%%%%%%%%%%%%%
In this paper, we design and implement an orbit feedback system based on deep reinforcement learning and address the following issues:
\textit{(i) System drifting;
(ii) Degenerated performance with traditional SVD-based linear method;
(iii) High dimensional control with reinforcement learning.}
Our contributions are summarized as follows:
\begin{enumerate}
\item In trajectory optimization, a model-based RL algorithm optimizes a policy neural network.
Trajectory optimization targets the entire control process instead of a single step.
Consequently, the trained policy chooses actions to ensure the stability of the future episode.
The trajectory sampling process simulates the control process to better fit the actual operation data of the machine.
On the other hand, the optimization runs on a differentiable surrogate model with the ideal environment setting (i.e., no noise).
This improves policy accuracy and accelerates the training process.
\item In online model optimization, the policy network is applied to the environment.
Real-time data is collected to train the system model adaptively.
Online model optimization targets adaptive control by interacting with the orbit feedback system.
This addresses the problem of system drifting.
The forward propagation neural network captures the non-linear behavior of the system with high accuracy.
Moreover, the training data for model optimization can be efficiently collected during beam daily operations. We do not need extra facility maintenance time for dataset collection.
\item We use the existing SVD-based method and the supervised learning model as the baseline and evaluate the model-based reinforcement learning system for the NSLS-II feedback control.
We compare their performance with the simulation environment.
Then we conduct real-world experiments in the NSLS-II feedback system for additional evaluations.
A neural network with three hidden layers of size 512 is trained to run on the NSLS-II feedback system, having 180 input and output dimensions.
Our method control stabilizes the RMS of beam position to \textasciitilde\SI{1}{\micro\metre}, about 80\% improvement compared to the current SVD-based method.
We plan to add our RL model to the production beam system and provide it to the operators of the NSLS-II storage ring.
\end{enumerate}

The remainder of this work is organized as follows.
\nameref{sec:related} section offers a short review of current machine-learning methods for storage rings.
\nameref{sec:background} section analyzes the orbit control challenges and explains the SVD-based algorithm and supervised learning model.
\nameref{sec:method} section details our feedback system based on reinforcement learning.
\nameref{sec:exp} section presents simulation results and experiment outcomes on the NSLS-II beam.
\nameref{sec:conclusion} section presents the conclusion and future plan.
%%%%%%%%%%%%%%%%%%%%%%%%%%%%%%%%%%%%%%%%%%%%%%%%%%%%%%%%%%%%%
%%%%%%%%%%%%%%%%%%%%%%%%%%%%%%%%%%%%%%%%%%%%%%%%%%%%%%%%%%%%%
\section{Related Work}\label{sec:related}
Deep learning and big-data-driven methods have drawn much attention recently.
The orbit feedback system is a multiple-input-multiple-output (MIMO) feedback system.
Treating the MIMO system as a black box, the neural network can model the inverse relationship between the machine status (inputs)  and the corrective actions (output) with supervised learning algorithms.  
In \cite{bozoki_neural_1994, fol_optics_2019, xiao_orbit_2019, dengjie_orbit_2021, schirmer_orbit_2020, ruichun_application_2021, chen_beam_2023}, neural networks were trained with supervised learning algorithm based on the input and output data.
The input and output dimensions were usually less than 100.
The network was trained with simulated data and validated with actual operating data for adapting to the real operation environment \cite{xiao_orbit_2019}.
In \cite{dengjie_orbit_2021}, the surrogate model was regressed from collected operating data, and a network was additionally trained with the surrogate model.
In \cite{schirmer_orbit_2020, ruichun_application_2021, chen_beam_2023}, real-time-control experiments were conducted on the storage ring to achieve low RMS errors or fast controls.

Reinforcement learning (RL) agents learn to make decisions by interacting with the environment.
Meier \cite{meier_orbit_2012} trained an actor-critic algorithm with input states and output actions of a small dimensionality (\(< 10\)) in a storage ring simulator to achieve real-time control.
Yang \cite{yang_online_2022} proposed a multi-agent DDPG design for orbit calibration in MEBT.
Apart from storage ring stabilization, studies \cite{kain_sample-efficient_2020, pang_autonomous_2020, wang_accelerator_2021, hirlaender_model-free_2022, scheinker_adaptive_2021, velotti_automatic_2022} also explored applying RL algorithm into other accelerator facilities, for example, linear accelerator, free-electron laser, etc.

Throughout the study, data for supervised learning is either generated from the simulation software or collected from historical operations.
This does not fit our situation for adaptive control.
Current studies on model-free RL only handle lower-dimension systems. However, our system has high dimensionality.
Thus, we design a model-based RL algorithm to achieve adaptive control with high dimensions.
%%%%%%%%%%%%%%%%%%%%%%%%%%%%%%%%%%%%%%%%%%%%%%%%%%%%%%%%%%%%%
%%%%%%%%%%%%%%%%%%%%%%%%%%%%%%%%%%%%%%%%%%%%%%%%%%%%%%%%%%%%%
\section{Background}\label{sec:background}
\subsection{Problem Definition}
\begin{figure}[htb]
\centering
\includegraphics[width=.9\linewidth]{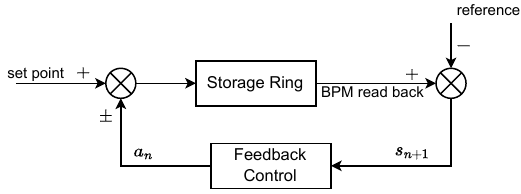}
\caption{An illustration of the closed loop feedback system.}\label{fig:closeloop}
\end{figure}

% The mathematical model for orbit feedback system can be approximated with Equation~\ref{eq:sys-model}.
The orbit feedback system runs in a closed control loop, shown in Fig.~\ref{fig:closeloop}.
The goal of the feedback controller is to produce \(a_n\), such that \(s_{n+1}\) maintains below the threshold.

We use the first-order approximation to model the feedback control system as follows: 
\begin{equation}\label{eq:sys-model}
s_{n+1} = s_{n} + R a_n,
\end{equation}
where \(a_n\) indicates the corrections applied, \(s_n\) and \(s_{n+1}\) are BPMs observed before and after we apply the correction, and \(R\) is the orbit response matrix.
%%%%%%%%%%%%%%%%%%%%%%%%%%%%%%%%%%%%%%%%%%%%%%%%%%%%%%%%%%%%%
\subsection{SVD-based Feedback Control}
\begin{figure}[htb]
\centerline{\includegraphics[width=.9\linewidth]{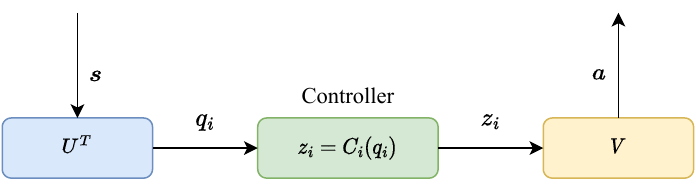}}
\caption{The SVD-based PID control: \(q_i\) and \(z_i\) stand for each component of the spectrum space after the transformation of the current input.}\label{fig:svd}
\end{figure}

We aim to control the next beam position \(s_{n+1}\) to 0.
A straightforward way is to solve for \(a_n = -R^{-1} s_n\). 
However, measurements of the ORM indicate the system has highly ill-posed dynamics.
 With measurement errors, the inverse of the response matrix could be extremely unstable.
Singular value decomposition (SVD) is used to inverse the problem \cite{corbett_algorithms_1994}.
In NSLS-II fast orbit feedback (FOFB) control, the SVD method combines with PID controller \cite{singh_nsls-ii_2015}.
Specifically, the controller is applied on each component of the spectrum space by doing SVD on the ORM: \(R = U \Sigma V^T\).
Figure~\ref{fig:svd} illustrates this process.

The parameter set for the PID controller is given by (with proportional component only)
\begin{equation}
z_i = C(q_i) = -\dfrac{\sigma_{i}}{\sigma_{i}^2 + \lambda} q_i,
\end{equation}
where \(\sigma_i\) is the \(i\)th singular value.

This process is proven identical to ridge regression.
\begin{equation}\label{eq:tiknov-reg}
\min_{a}\ \norm{R a + s}_2^2 + \lambda \norm{a}_2^2.
\end{equation}
%%%%%%%%%%%%%%%%%%%%%%%%%%%%%%%%%%%%%%%%%%%%%%%%%%%%%%%%%%%%%
\subsection{Supervised Learning Model}\label{sec:method-sup}
Supervised learning learns from labeled data.
Several research efforts applied supervised learning to the orbit feedback control problems \cite{bozoki_neural_1994, fol_optics_2019, xiao_orbit_2019, dengjie_orbit_2021, schirmer_orbit_2020, ruichun_application_2021, chen_beam_2023}.
For our problem, we train a neural network to generate action \(a_n\) given  current state \(s_n\) as input.
In the following context, we denote this network \(\pi(s_n)\).

Given the dataset \(\mathcal{D} \in R^{m}\times R^{m}\), the loss function for supervised learning is
\begin{equation}
L(w_\pi) = \mathop{\mathbb{E}}_{(s_n, a_n) \in \mathcal{D}}\norm{a_n - \pi(s_n)}.    
\end{equation}

\subsubsection{Dataset preparation}
The training dataset \(\mathcal{D}\) can be obtained by:
\textit{(i) extracting state-action pairs directly from the data archive of the running machine;
(ii)\label{pt:sup-svd} running simulation software to generate states randomly and using an SVD-based algorithm to produce the corresponding action;
(iii) running forward simulations to generate random actions as inputs and produce the subsequent states.} 

% \subsubsection{Other types of loss}
% Assume we are using method (ii) for data collection, we can rewrite the loss function by multiplying \(-R\).
% \[L(w_\pi) = \mathop{\mathbb{E}}_{s_n \in \mathcal{N}}\norm{s_n + R\pi(s_n)}.\]

% To avoid overfitting, we can add regularization to the action, inspired by Equation~\ref{eq:tiknov-reg}
% \begin{equation}\label{eq:sup-loss-reg}
% L(w_\pi) = \mathop{\mathbb{E}}_{s_n \in \mathcal{N}}\left[\norm{s_n + R\pi(s_n)} + \lambda \norm{\pi(s_n)}\right].
% \end{equation}
%%%%%%%%%%%%%%%%%%%%%%%%%%%%%%%%%%%%%%%%%%%%%%%%%%%%%%%%%%%%%
\subsection{Reinforcement Learning Framework}\label{sec:method-rl}
RL aims to obtain a strategy to maximize the expected cumulative returns by interacting with the system.
A typical RL framework comprises the tuple $(S, A, r, P, \gamma)$.
State space \(S\) describes all possible running statuses of the storage ring, and action space \(A\) specifies the action to alter the system.
Then the system can be abstracted as the probability mapping of the next state given the current state and action, say \(P(s_{n+1}|s_n, a_n)\).
Given a reward function \(r(s, a)\), we aim to find the optimal control \(a_n = \pi(s_n)\), called policy function, which maximizes the expected total reward
\begin{equation}\label{eq:rew}
R(\pi) = \mathop{\mathbb{E}}_{s_0} \left[\sum_n \gamma^n r(s_n, a_n)\right]
\end{equation}
over the whole trajectory.
Here \(\gamma\) is the decay parameter to ensure the convergence of the expectation.

For the beam control problem in NSLS-II, we take the current BPMs as \(S\), and the control signal as action \(A\).
If we do not involve noises and machine drifting, the system model is deterministic and given by Eq.~(\ref{eq:sys-model}).
%%%%%%%%%%%%%%%%%%%%%%%%%%%%%%%%%%%%%%%%%%%%%%%%%%%%%%%%%%%%%
%%%%%%%%%%%%%%%%%%%%%%%%%%%%%%%%%%%%%%%%%%%%%%%%%%%%%%%%%%%%%
\section{Method}\label{sec:method}
% There are several works using reinforcement learning to control the accelerator particle beam in multiple systems \cite{kain_sample-efficient_2020, pang_autonomous_2020, wang_accelerator_2021, hirlaender_model-free_2022, scheinker_adaptive_2021, velotti_automatic_2022}.
% In \cite{kain_sample-efficient_2020}, a NAF agent was successfully trained online, with up to 17 observations and 16 control dimensions.

% These preliminary studies demonstrate the feasibility of RL for physical control problems.
% However, these systems are different from the beam in the storage ring, and there are no current attempts to apply RL method to stabilize the storage ring.
% In this section, we introduce the RL algorithms into the beam control problem in NSLS-II storage ring.

% \subsection{Model-based RL Algorithms}
Model-based RL algorithms optimize the expected total reward based on the system model information.
This section explores a model-based way to optimize the policy network.

\begin{figure}[htb]
\centerline{\includegraphics[width=\linewidth]{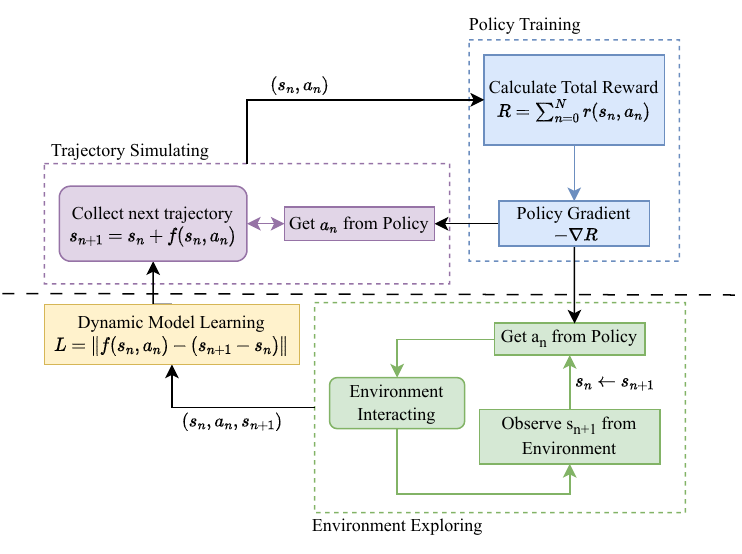}}
\caption{The data flow for model-based RL with online model optimization.}\label{fig:mb-diag}
\end{figure}

Figure~\ref{fig:mb-diag} shows the entire process for our framework when we run the algorithm online. The upper part of Figure~\ref{fig:mb-diag} presents the trajectory optimization while the lower is for online model optimization.

\subsection{Trajectory Optimization}
The system model is given by Eq.~(\ref{eq:sys-model}), which is a differentiable model.
Therefore, we can always generate a differentiable reward function when running the policy on this model.
Leveraging the \textit{autograd} engine in PyTorch, we do not have to bother calculating the complex gradient by hand, but collect the gradient information directly from the trajectory sampling process.

We utilize policy gradient directly to train the policy network.
In the trajectory sampling process, we first sample a random state \(s_0\), then iteratively calculate the new state for time horizon \(N\).
\begin{equation}
    s_{n+1} = s_n + R \pi(s_n),\quad\text{\(n\) from 0 to \(N-1\)}.
\end{equation}
Then the loss function will be the negative of the total reward
\begin{equation}
    L(\pi) = -\sum_{i=0}^{N-1} r(s_i, \pi(s_i)).
\end{equation}

The policy network will be updated based on \(\nabla_{\pi} L\). Details are shown in Algorithm~\ref{alg:mb}.
\begin{algorithm}[htb]
\caption{Policy Gradient with Trajectory Sampling}\label{alg:mb}
\begin{algorithmic}
\Require The system model \(R\), Reward function \(r(s, a)\)
\Ensure Policy \(\pi(s)\)
\State Initialize neural network \(\pi\).
\While{total episodes less than limit}
Initialize \(s_0\).
\While{steps less than limit N}
\State \(a_n \leftarrow \pi(s_n)\).
\State \(s_{n+1} = s_n + R \pi(s_n)\).
\State Save \(s_n, s_{n+1}, a_n\) for training.
\State \(s_n \leftarrow s_{n+1}\).
\EndWhile
\State Calculate policy loss based on expected reward.
\[L(w_\pi) = -\sum_{i=0}^{N-1} r(s_i, a_i).\]
\State Update weight \(w_\pi\) based on the \(\nabla L_{w_\pi}\).
\EndWhile
\end{algorithmic}
\end{algorithm}
%%%%%%%%%%%%%%%%%%%%%%%%%%%%%%%%%%%%%%%%%%%%%%%%%%%%%%%%%%%%%
\subsection{Online Model Optimization}
Trajectory optimization pre-trains the policy network using the given system model~(\ref{eq:sys-model}) with \(R\).
However, this \(R\) might not be accurate and not represent the actual system behavior.
Thus, after interacting with the environment, we can update the system model based on the collected data.

To fit the latest system model, data collection should not happen in Algorithm~\ref{alg:mb}.
Instead, the policy should run in parallel on the physical machine to collect the data point \((s_n, a_n, s_{n+1}) \in \mathcal{D}\).
Then we can fit a new response matrix through least square.
\begin{equation}
\hat{R} = \argmin_R \mathop{\mathbb{E}}_{\mathcal{D}} \norm{(s_{n+1} - s_n) - R a_n}_2^2.
\end{equation}
The resulting \(\hat{R}\) is then used to replace the original \(R\) matrix in Algorithm~\ref{alg:mb}.

Furthermore, the system model for online training does not have to be a linear function.
In fact, the real machine does not have linear dynamics.
Thus, we train another neural network \(f(s_n, a_n)\) for forward system model learning.
This network is trained in a supervised learning way.
\begin{equation}
L(w_f) = \norm{(s_{n+1} - s_n) - f(s_n, a_n)}.    
\end{equation}
The system model then becomes
\begin{equation}
s_{n+1} = s_n + f(s_n, a_n).    
\end{equation}
\(f(s_n, a_n)\) replaces the original system model in Algorithm~\ref{alg:mb} for further training.
%%%%%%%%%%%%%%%%%%%%%%%%%%%%%%%%%%%%%%%%%%%%%%%%%%%%%%%%%%%%%
%%%%%%%%%%%%%%%%%%%%%%%%%%%%%%%%%%%%%%%%%%%%%%%%%%%%%%%%%%%%%
\section{Experiments}\label{sec:exp}
We experiment on the simulation environment with SVD-based, supervised learning, and our model-based RL methods.
The trained models are tested in the NSLS-II storage ring for further validation.
%%%%%%%%%%%%%%%%%%%%%%%%%%%%%%%%%%%%%%%%%%%%%%%%%%%%%%%%%%%%%
\subsection{Experiment Setups}
\subsubsection{Environment Setup}
In preliminary experiments, we employ a simulated environment, which runs the system model [Eq.~(\ref{eq:sys-model})] to produce the next state.
The input dimension and output dimension are both 180.
However, some features are added to address two key properties of the actual machine:
\textit{(i) two ORMs are measured at different machine states. The ORM used for system model will drift from one to another over iterations;
(ii) observation noises are added to the BPM readback to simulate electronic noise.}
\subsubsection{Evaluation Metrics}
For each algorithm, we run \(N\) long trajectories with length \(m\) in the simulated environment.
We calculate the root mean square for each trajectory and plot the average of the RMS with its variance across \(N\) trajectories.
The following performance metrics are considered:
\textit{(i) best state RMS; (ii) worst state RMS; (iii) final state RMS; (iv) steps needed to reduce the RMS to a certain threshold.}
%%%%%%%%%%%%%%%%%%%%%%%%%%%%%%%%%%%%%%%%%%%%%%%%%%%%%%%%%%%%%
\subsubsection{Neural network details}\label{sec:nn-design}
For this problem, we will use a deep neural network with a 180-dimension input and output layer, and 3 hidden layers of 512 dimensions each. This gives us approximately 0.7\,M parameters in total.

This particular problem has a special property of the action taken.
That is to take zero action if the state is already zero.
To fit this property, we will use unbiased linear layers (set \(b_i\) to zero), and use the 
hyperbolic tangent function (\(\tanh\)) as the activation function.

We use Adam \cite{kingma_adam_2017} as the training algorithm with learning rate \(10^{-4}\).
This algorithm is considered to achieve superior performance in machine learning research.
The same policy network design, training algorithm, and initial parameters will be used consistently throughout the experiment.
%%%%%%%%%%%%%%%%%%%%%%%%%%%%%%%%%%%%%%%%%%%%%%%%%%%%%%%%%%%%%
\subsection{Simulation Experiment}

% \begin{figure*}[htb]
% % \captionsetup{width=.8\textwidth}
% \centerline{\includegraphics[width=.8\textwidth]{matrices.pdf}}
% \caption{Visualization of different inverse systems. Left to right: the original orbit response matrix; SVD-based persudo inverse; identity response of NN model trained from supervised learning; identity response of NN model from model-based RL (at final state).}\label{fig:inverse-matrix}
% \end{figure*}
We evaluate three methods with identical environment settings: the SVD-based, supervised learning, and our method.
For our method, the policy network is pre-trained with trajectory optimization before interacting with the environment.
The reward function used is the negative of the RMS.
% plus the regularization term of action RMS to make the training stable.
We sample 1,000 trajectories with a length of 10,000 for simulation, and plot the mean and std across different runs.

% Figure~\ref{fig:inverse-matrix} shows all the matrix representations of the inverse systems for different algorithms after training.
% All the inverse matrix is near diagonal.
% This has the physical meaning that the correction response should only depend on the BPM locally.
% The supervised learning and model-based RL ones have some tiny structures inside.
% This could indicate that these models are overfitting.
% However, the model-based RL agent always adapts to the latest system dynamic.
% Thus overfitting is fine here.

\begin{figure}[htb]
% \captionsetup{width=.7\textwidth}
\centerline{\includegraphics[width=\linewidth]{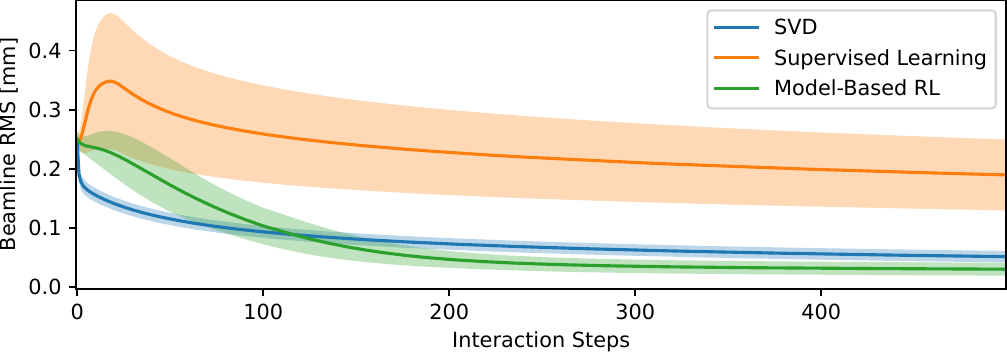}}
\caption{Plot of trajectory at first 500 interactions. The solid line and shadows show the mean and standard deviation across different simulations.}\label{fig:plot-short}
\end{figure}

Figure~\ref{fig:plot-short} displays a short trajectory period, illustrating how our method is capable of recovering from poor machine BPMs.
Our findings indicate that supervised learning approaches are unable to reduce the state to a smaller RMS value.
This is likely due to overfitting so that the network cannot adapt to drifting environments. 
Our method initially converges slower than the SVD method.
Interactions with the environment allow the agent to acquire new information about changes in environment dynamics, and refine its policies accordingly.
Eventually, our method continues to reduce the RMS and surpasses the SVD method.

\begin{figure}[htb]
% \captionsetup{width=.7\textwidth}
\centerline{\includegraphics[width=\linewidth]{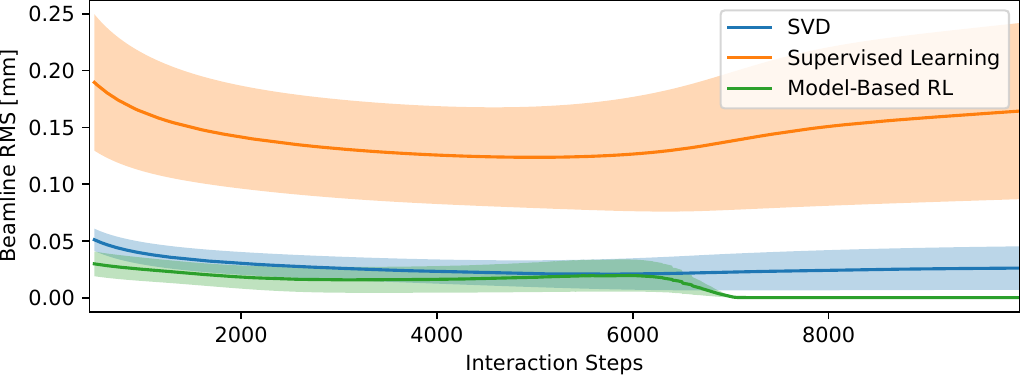}}
\caption{Plot of trajectory for long term run (10000 steps).}\label{fig:plot-long}
\end{figure}

Figure~\ref{fig:plot-long} simulates the long-run experiments.
As the system's dynamics change over time, other methods cannot capture this and result in degenerated performance.
For our method, the longer it interacts with the environment, the more robust it will be.
At the final iteration, our agent controls the beam RMS down to the machine measurement accuracy (\textasciitilde\SI{0.2}{\micro\metre}).
Table~\ref{tab:summary} summarizes the key metric in the experiment.
\begin{table*}[htb]
\centering
\begin{threeparttable}
\caption{Summary of the Experiment Result}\label{tab:summary}
\begin{tabular}{llllr}
\toprule
{} &   \textbf{Min RMS} &   \textbf{Max RMS} &  \textbf{Final RMS} &  \textbf{Steps to reach 0.05} \\
\midrule
SVD-Based Method    &  0.021 &  0.25\sym{*} &   0.026 &           532 \\
Supervised Learning &  0.12 &  0.35 &   0.16 &             N/A \\
Model-Based RL      &  \textbf{0.00028} &  0.25\sym{*} &   \textbf{0.00033} &           \textbf{187} \\
\bottomrule
\end{tabular}
\smallskip\footnotesize
\begin{tablenotes}
\item[*]Initial state
\end{tablenotes}
\end{threeparttable}
\end{table*}
%%%%%%%%%%%%%%%%%%%%%%%%%%%%%%%%%%%%%%%%%%%%%%%%%%%%%%%%%%%%%
%%%%%%%%%%%%%%%%%%%%%%%%%%%%%%%%%%%%%%%%%%%%%%%%%%%%%%%%%%%%%
\subsection{Experiments on NSLS-II System}
\begin{figure*}[htbp!]
\centerline{\includegraphics[width=.7\textwidth]{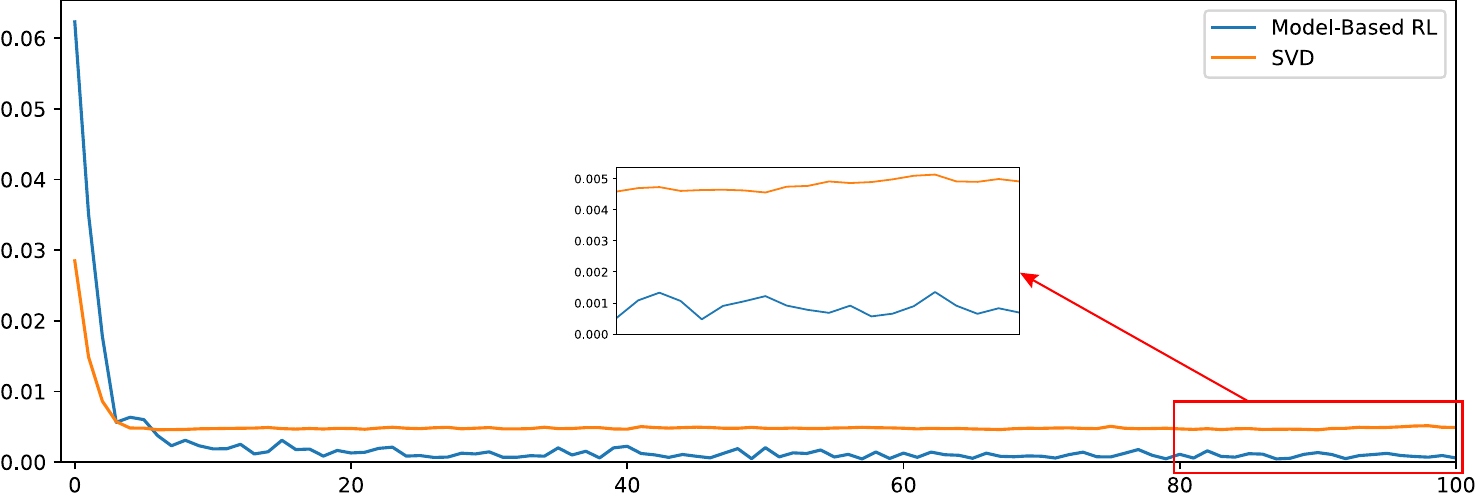}}
\caption{Performance of Model-Based RL Algorithm on NSLS-II Storage Ring.}\label{fig:nsls-mb}
\end{figure*}

Based on the above work, we tested our machine-learning method directly on the storage ring beam.
Due to the study time limit, we could not do a long-run test for our method.
The online model optimization was performed once to fit into the current machine status.
The results are shown in Fig.~\ref{fig:nsls-mb}.

In the experiment, we randomly kicked the beam off its original position and applied our models to control the orbit back.
In Fig.~\ref{fig:nsls-mb}, the SVD-based method only stabilizes the beam RMS to \textasciitilde\SI{5}{\micro\metre}, while our method reaches \textasciitilde\SI{1}{\micro\metre}, showing a 80\% improvement of RMS values.
%%%%%%%%%%%%%%%%%%%%%%%%%%%%%%%%%%%%%%%%%%%%%%%%%%%%%%%%%%%%%\norm{
%%%%%%%%%%%%%%%%%%%%%%%%%%%%%%%%%%%%%%%%%%%%%%%%%%%%%%%%%%%%%
\section{Conclusion}\label{sec:conclusion}
This study investigates the machine-learning methods for controlling the beam in the NSLS-II storage ring.
The feedback system of NSLS-II is modeled by the orbit response matrix.
The ORM cannot be obtained precisely due to machine internal status drift, environmental noise, and non-linear behavior of the system.
Thus, the SVD controller leads to beam drift over time.
Supervised learning is unsuitable for our control system because it tends to overfit and is not adaptive to machine drift.
The model-based RL algorithm runs interactively with the environment to achieve adaptive control.   
Trajectory optimization optimizes the expected total reward using policy gradient.
This approach involves using a neural network to learn the non-linear dynamics of the beam orbit system and extracting the optimal control signal over the trajectory.
Online model optimization adaptively fits the current environment behavior by collecting real-time running data of the policy.
Through both simulation and real-world experiments, our proposed method outperforms many existing algorithms and achieves 80\% improvement compared to the current SVD-based method deployed at NSLS-II.

% Although we already conduct successful machine-learning experiments on the NSLS-II system.
% There are still many remaining works to do.
% For the model-based algorithm, we only verified part of its functionality but did not test for online model learning, and how it will improve the control performance.
% In high-noise physical environments, the algorithm may behave very differently.
% Thus, we need to perform further simulation, conduct future experiments in online environments, and update the algorithm that fit for high noise environment.

The adaptive control for our method runs in an overfitting way.
That means we tried to consume as much training time as to keep track of the system drift.
It could lead to a biased dataset for the algorithm to train on, ultimately impeding the algorithm's ability to learn and generalize effectively.
To mitigate this issue, we recommend collecting a significant amount of data before starting model optimization.
We propose exploring new algorithms that allow the system to detect performance degeneration in real-time and perform online model optimization on demand.
\section*{Acknowledgements}
This material is based upon work supported by the U.S. Department of Energy, Office of Science, Office of Advanced Scientific Computing Research, Office of Nuclear Physics, under Award Number DE-SC0019518.

\end{document}